\documentstyle[aasms4]{article}
\makeatletter
  
  \@addtoreset{equation}{section}
\makeatother

\begin{document}

\title{RELATIVISTIC CORRECTIONS TO THE SUNYAEV-ZEL'DOVICH EFFECT FOR CLUSTERS OF GALAXIES}

\author{NAOKI ITOH\altaffilmark{1}}

\affil{Department of Physics, Sophia University, 7-1 Kioi-cho, Chiyoda-ku, Tokyo, 102, Japan}

\author{YASUHARU KOHYAMA\altaffilmark{2}}

\affil{Fuji Research Institute Corporation, 2-3 Kanda-Nishiki-cho, Chiyoda-ku, Tokyo, 101, Japan}

\centerline{AND}

\author{SATOSHI NOZAWA\altaffilmark{3}}

\affil{Josai Junior College for Women, 1-1 Keyakidai, Sakado-shi, Saitama, 350-02, Japan}

\altaffiltext{1}{n\_itoh@hoffman.cc.sophia.ac.jp}
\altaffiltext{2}{kohyama@crab.fuji-ric.co.jp}
\altaffiltext{3}{snozawa@venus.josai.ac.jp}

\begin{abstract}

  We study the generalized Kompaneets equation (kinetic equation for the photon distribution function taking into account the Compton scattering by electrons) using a relativistically covariant formalism.  Using the generalized Kompaneets equation, we derive an analytic expression for the Sunyaev-Zel'dovich effect, which takes into account up to $O(\theta_{e}^{5})$ terms, where $\theta_{e}=k_{B}T_{e}/mc^{2}$ is the relativistic expansion parameter and $T_{e}$ is the electron temperature.  We carefully study the applicable region of the obtained analytic expression  by comparing with the result of the direct numerical integration.  We conclude that the present analytic expression can be reliably applied to the calculation of the Sunyaev-Zel'dovich effect for $k_{B}T_{e} \leq 15{\rm keV}$, which is the highest electron temperature in the presently known clusters of galaxies.  Therefore, the present analytic expression can be applied to all known clusters of galaxies.

\end{abstract}


\section{INTRODUCTION}

  Compton scattering of the cosmic microwave background (CMB) radiation by hot intracluster gas --- the Sunyaev-Zel'dovich effect --- (Zel'dovich \& Sunyaev 1969; Sunyaev \& Zel'dovich 1972, 1980) provides a useful method to measure the Hubble constant $H_{0}$ (Gunn 1978; Silk \& White 1978; Birkinshaw 1979; Cavaliere, Danese, \& De Zotti 1979; Birkinshaw, Hughes, \& Arnaud 1991; Birkinshaw \& Hughes 1994; Myers et al. 1995; Herbig et al. 1995; Jones 1995; Markevitch et al. 1996; Holzapfel et al. 1997).  The Sunyaev-Zel'dovich formula has been derived from a kinetic equation for the photon distribution function taking into account the Compton scattering by electrons --- the Kompaneets equation --- (Kompaneets 1957; Weymann 1965).  The Kompaneets equation has been derived with the non-relativistic approximation for the electron.  However, the electrons in the clusters of galaxies are extremely hot, $k_{B} T_{e}$ = $5 \sim 15$keV (Arnaud et al. 1994; Markevitch et al. 1994; Markevitch et al. 1996; Holzapfel et al. 1997).

  Recently attempts have been made to include the relativistic corrections in the Sunyaev-Zel'dovich effect (Rephaeli 1995; Rephaeli \& Yankovitch 1997).  However, it appears that the calculations have not been carried out in a manifestly covariant form.  For example, equation (4) in Rephaeli (1995), which comes from Chandrasekhar (1950), is a non-relativistic formula.  Since the extension of the Kompaneets equation to the relativistic regime is extremely important in view of many recent measurements of the Hubble constant $H_{0}$ with the use of the Sunyaev-Zel'dovich effect, we will solve the kinetic equation for the photon distribution function in a manifestly covariant form taking into account the Compton scattering by electrons.

  Very recently a generalized Kompaneets equation has been derived by two groups (Stebbins (1997); Challinor \& Lasenby (1997)).  By using the generalized Kompaneets equation, analytic expressions for the Sunyaev-Zel'dovich effect have been derived as a power series of $\theta_{e}$ = $k_{B}T_{e}/mc^{2}$, where $T_{e}$ and $m$ are the electron temperature and the electron mass, respectively.  It has been shown that the results obtained by a power series expansion agree with the previous numerical calculations by Rephaeli (1995) and Rephaeli \& Yankovitch (1997).  Analytic expressions are compact and extremely useful to study the Sunyaev-Zel'dovich effect.  On the other hand, it has been pointed out by Challinor \& Lasenby (1997) that the convergence of the power series expansion in $\theta_{e}$ is slow.  They suspect that the expansion is an asymptotic expansion.  Therefore, it is extremely important to study the extent of the $\theta_{e}$ region where the analytic expressions can be applied.  This is the main subject in the present paper.  Following the approach of Challinor \& Lasenby (1997), we will derive analytic expressions for the intensity change of the photon spectrum.  In the derivation we will take into account the relativistic terms up to the fifth order in $\theta_{e}$.  In order to examine the accuracy of the analytic expressions derived from the power series expansion, we will directly integrate the Boltzmann equation numerically.  Comparing these results we will  carefully study the valid region for the electron temperature $T_{e}$ and for the photon angular frequency $\omega$ where the present analytic expressions can be reliably applied.

  The present paper is organized as follows.  Extension of the Kompaneets equation to the relativistic regime will be treated and an analytic expression for the intensity change of the photon spectrum will be derived in $\S$ 2.  In $\S$ 3 we will study the accuracy of the analytic expressions  by comparing with the numerical results. Concluding remarks will be given in $\S$ 4.

\section{GENERALIZED KOMPANEETS EQUATION}

  In this section we will extend the Kompaneets equation to the relativistic regime.  We will formulate the kinetic equation for the photon distribution function using a relativistically covariant formalism (Berestetskii, Lifshitz, \& Pitaevskii 1982; Buchler \& Yueh (1976).  As a reference system, we choose the system which is fixed to the center of mass of the cluster of galaxies.  This choice of the reference system affords us to carry out all the calculations in the most straightforward way.  We will use the invariant amplitude for the Compton scattering as given by Berestetskii, Lifshitz, \& Pitaevskii (1982) and by Buchler \& Yueh (1976).

 The time evolution of the photon distribution function $n(\omega)$ is written as 
\begin{eqnarray}
\frac{\partial n(\omega)}{\partial t} & = & -2 \int \frac{d^{3}p}{(2\pi)^{3}} d^{3}p^{\prime} d^{3}k^{\prime} \, W \,
\left\{ n(\omega)[1 + n(\omega^{\prime})] f(E) - n(\omega^{\prime})[1 + n(\omega)] f(E^{\prime}) \right\} \, ,  \\
W & = & \frac{(e^{2}/4\pi)^{2} \, \overline{X} \, \delta^{4}(p+k-p^{\prime}-k^{\prime})}{2 \omega \omega^{\prime} E E^{\prime}} \, ,  \\
\overline{X} & = & - \left( \frac{\kappa}{\kappa^{\prime}} + \frac{\kappa^{\prime}}{\kappa} \right) + 4 m^{4} \left( \frac{1}{\kappa} + \frac{1}{\kappa^{\prime}} \right)^{2} 
 - 4 m^{2} \left( \frac{1}{\kappa} + \frac{1}{\kappa^{\prime}} \right) \, ,  \\
\kappa & = & - 2 (p \cdot k) \, = \, - 2 \omega E \left( 1 - \frac{\mid \vec{p} \mid}{E} {\rm cos} \alpha \right) \, ,  \\
\kappa^{\prime} & = &  2 (p \cdot k^{\prime}) \, = \, 2 \omega^{\prime} E \left( 1 - \frac{\mid \vec{p} \mid}{E} {\rm cos} \alpha^{\prime} \right) \, .
\end{eqnarray}
In the above $W$ is the transition probability corresponding to the Compton scattering.  The four-momenta of the initial electron and photon are $p = (E, \vec{p})$ and $k = (\omega, \vec{k})$, respectively.  The four-momenta of the final electron and photon are $p^{\prime} = (E^{\prime}, \vec{p}^{\prime})$ and $k^{\prime} = (\omega^{\prime}, \vec{k}^{\prime})$, respectively.  The angles $\alpha$ and $\alpha^{\prime}$ are the angles between $\vec{p}$ and $\vec{k}$, and between $\vec{p}$ and $\vec{k}^{\prime}$, respectively.  Throughout this paper, we use the natural unit $\hbar = c = 1$ unit, unless otherwise stated explicitly.

  By ignoring the degeneracy effects, we have the relativistic Maxwellian distribution for electrons with temperature $T_{e}$ as follows
\begin{eqnarray}
f(E) & = & \left[ e^{\left\{(E - m)-(\mu - m) \right\}/k_{B}T_{e}} \, + \, 1 \right]^{-1}  \nonumber \\
& \approx & e^{-\left\{K-(\mu - m)\right\}/k_{B}T_{e}} \, ,
\end{eqnarray}
where $K \equiv (E - m)$ is the kinetic energy of the initial electron, and $(\mu - m)$ is the non-relativistic chemical potential of the electron. 
We now introduce the quantities
\begin{eqnarray}
x &  \equiv &  \frac{\omega}{k_{B}T_{e}}  \, ,  \\
\Delta x &  \equiv &  \frac{\omega^{\prime} - \omega}{k_{B}T_{e}}  \, .
\end{eqnarray}
Substituting eqs.\ (2.6) -- (2.8) into eq.\ (2.1), we obtain
\begin{equation}
\frac{\partial n(\omega)}{\partial t} =  -2 \int \frac{d^{3}p}{(2\pi)^{3}} d^{3}p^{\prime} d^{3}k^{\prime} \, W \, f(E) \,
\left\{ [1 + n(\omega^{\prime})] n(\omega) -  [1 + n(\omega)] n(\omega^{\prime}) e^{ \Delta x }  \right\} \, .
\end{equation}
Eq.\ (2.9) is our basic equation.  One can numerically integrate this 
equation directly.  We will perform this integration in $\S$3.

  Following Challinor \& Lasenby (1997), we expand eq.\ (2.9) in powers of 
$\Delta x$ by assuming $\Delta x  \, \ll 1$.  We obtain the Fokker-Planck expansion
\begin{eqnarray}
\frac{ \partial n(\omega)}{ \partial t} & = & 
 2 \left[ \frac{ \partial n}{ \partial x} + n(1+n) \right] \, I_{1} 
  \nonumber  \\
& + & 2 \left[ \frac{ \partial^{2} n}{ \partial x^{2}} 
+ 2(1+n) \frac{ \partial n}{ \partial x} + n(1+n)  \right] \, I_{2}
  \nonumber  \\
& + & 2 \left[\frac{ \partial^{3} n}{ \partial x^{3}} 
+ 3(1+n) \frac{ \partial^{2} n}{ \partial x^{2}} 
+ 3(1+n) \frac{ \partial n}{ \partial x} + n(1+n)  \right] \, I_{3}
  \nonumber \\
& + & 2 \left[\frac{ \partial^{4} n}{ \partial x^{4}} 
+ 4(1+n) \frac{ \partial^{3} n}{ \partial x^{3}}
+ 6(1+n) \frac{ \partial^{2} n}{ \partial x^{2}}
+ 4(1+n) \frac{ \partial n}{ \partial x} + n(1+n)  \right] \, I_{4}
  \nonumber \\
& + & \cdot \cdot \cdot  \, \, \, ,
\end{eqnarray}
where
\begin{equation}
I_{k} \, \equiv \,  \frac{1}{k !} \int \frac{d^{3}p}{(2\pi)^{3}} d^{3}p^{\prime} d^{3}k^{\prime} \, W \, f(E) \, (\Delta x)^{k} \, .
\end{equation}
Analytic integration of eq.\ (2.11) is not possible except for doing power series expansions of the integrand.  Technically speaking, there are several choices for the expansion parameter of the integrand of eq.\ (2.11). They are, for example, $p/m$, $K/m \equiv (E/m - 1)$, and $v \equiv p/E$. 
It is important to note that obtained analytic expressions of $I_{k}$ after the integration do not depend on the choice of the expansion parameter. It is also extremely important to note that the expansions in terms of these variables are {\it asymptotic expansions} in $I_{k}$.  Therefore, not only the convergence is very slow but also the accuracy of the analytic expressions have to be carefully examined for the parameter region considered.  This is one of our main subjects in the present paper.

  Challinor \& Lasenby (1997) carried out a calculation up to $O(\theta_{e}^{3})$ terms.  We will carry out a calculation up to $O(\theta_{e}^{5})$ terms in the present paper.  The calculation of $I_{k}$ has been performed with a symbolic manipulation computer algebra package {\it Mathematica}.  We obtain
\begin{eqnarray}
I_{1} & = & \frac{1}{2} \sigma_{T} N_{e} \theta_{e} x \, \left\{ \, 4 - x 
\, + \, \theta_{e} \left(10 - \frac{47}{2} x + \frac{21}{5} x^{2} \right) 
\right. \nonumber \\
& + & \theta_{e}^{2} \left( \frac{15}{2} - \frac{1023}{8} x + \frac{567}{5} x^{2} - \frac{147}{10} x^{3} \right)  \nonumber  \\
 & + & \theta_{e}^{3} \left( - \frac{15}{2} - \frac{2505}{8} x + \frac{9891}{10} x^{2} - \frac{9551}{20} x^{3} + \frac{1616}{35} x^{4} \right)  \nonumber  \\
 & + & \left. \theta_{e}^{4} \left( \frac{135}{32} - \frac{30375}{128} x + \frac{177849}{40} x^{2} - \frac{472349}{80} x^{3} + \frac{63456}{35} x^{4} - \frac{940}{7} x^{5} \right)  \right\}  \, ,   \\
         \nonumber  \\
I_{2} & = & \frac{1}{2} \sigma_{T} N_{e} \theta_{e} x^{2} \, \left\{ \, 1  
\, + \, \theta_{e} \left(\frac{47}{2} - \frac{63}{5} x + \frac{7}{10} x^{2} \right)
  \right.  \nonumber \\ 
& + & \theta_{e}^{2} \left( \frac{1023}{8} - \frac{1302}{5} x + \frac{161}{2} x^{2} - \frac{22}{5} x^{3} \right)  \nonumber  \\
 & + & \theta_{e}^{3} \left( \frac{2505}{8} - \frac{10647}{5} x + \frac{38057}{20} x^{2} - \frac{2829}{7} x^{3} + \frac{682}{35} x^{4} \right)  \nonumber  \\
 & + & \left. \theta_{e}^{4} \left( \frac{30375}{128} - \frac{187173}{20} x + \frac{1701803}{80} x^{2} - \frac{44769}{4} x^{3} + \frac{61512}{35} x^{4} - \frac{510}{7} x^{5} \right)  \right\}  \, ,   \\
         \nonumber  \\
I_{3} & = & \frac{1}{2} \sigma_{T} N_{e} \theta_{e} x^{3} \, \left\{  \,
\theta_{e} \left(\frac{42}{5} - \frac{7}{5} x \right)  \right. \nonumber \\
& + & \theta_{e}^{2} \left( \frac{868}{5} - \frac{658}{5} x + \frac{88}{5} x^{2} - \frac{11}{30} x^{3} \right)  \nonumber  \\
 & + & \theta_{e}^{3} \left( \frac{7098}{5} - \frac{14253}{5} x + \frac{8084}{7} x^{2} - \frac{3503}{28} x^{3} + \frac{64}{21} x^{4} \right)  \nonumber  \\
 & + & \left. \theta_{e}^{4} \left( \frac{62391}{10} - \frac{614727}{20} x + 28193 x^{2} - \frac{123083}{16} x^{3} + \frac{14404}{21} x^{4} - \frac{344}{21} x^{5} \right)  \right\}  
\, ,        \\
         \nonumber  \\
I_{4} & = & \frac{1}{2} \sigma_{T} N_{e} \theta_{e} x^{4} \, \left\{ \, \frac{7}{10} \theta_{e}  \right. \nonumber  \\
& + & \theta_{e}^{2} \left( \frac{329}{5} - 22 x + \frac{11}{10} x^{2} \right) 
\nonumber \\
& + & \theta_{e}^{3} \left( \frac{14253}{10} - \frac{9297}{7} x + \frac{7781}{28} x^{2} - \frac{320}{21} x^{3} + \frac{16}{105} x^{4} \right)  \nonumber  \\
 & + & \left. \theta_{e}^{4} \left( \frac{614727}{40} - \frac{124389}{4} x + \frac{239393}{16} x^{2} - \frac{7010}{3} x^{3} + \frac{12676}{105} x^{4} - \frac{11}{7} x^{5} \right)  \right\}  \, ,   \\
         \nonumber  \\
I_{5} & = & \frac{1}{2} \sigma_{T} N_{e} \theta_{e} x^{5} \, \left\{  \,
\theta_{e}^{2} \left(\frac{44}{5} - \frac{11}{10} x \right)  \right.  \nonumber  \\
& + & \theta_{e}^{3} \left( \frac{18594}{35} - \frac{36177}{140} x + \frac{192}{7} x^{2} - \frac{64}{105} x^{3} \right)  \nonumber  \\
 & + & \left. \theta_{e}^{4} \left( \frac{124389}{10} - \frac{1067109}{80} x + 3696 x^{2} - \frac{5032}{15} x^{3} + \frac{66}{7} x^{4} - \frac{11}{210} x^{5} \right)  \right\}  \, , \\
         \nonumber  \\
I_{6} & = & \frac{1}{2} \sigma_{T} N_{e} \theta_{e} x^{6} \, \left\{ \, \frac{11}{30} \theta_{e}^{2}  \right.  \nonumber   \\
& + & \theta_{e}^{3} \left( \frac{12059}{140} - \frac{64}{3} x + \frac{32}{35} x^{2} \right) \nonumber \\
& + & \left. \theta_{e}^{4} \left( \frac{355703}{80} - \frac{8284}{3} x + \frac{6688}{15} x^{2} - 22 x^{3} + \frac{11}{42} x^{4} \right)  \right\}  \, ,   \\
         \nonumber  \\
I_{7} & = & \frac{1}{2} \sigma_{T} N_{e} \theta_{e} x^{7} \, \left\{  \,
\theta_{e}^{3} \left(\frac{128}{21} - \frac{64}{105} x \right)  \right.  \nonumber  \\
& + & \left. \theta_{e}^{4} \left( \frac{16568}{21} - \frac{30064}{105} x + \frac{176}{7} x^{2} - \frac{11}{21} x^{3} \right)   \right\}  \, , \\
         \nonumber  \\
I_{8} & = & \frac{1}{2} \sigma_{T} N_{e} \theta_{e} x^{8} \, \left\{ \, \frac{16}{105} \theta_{e}^{3}  \, + \, \theta_{e}^{4} \left( \frac{7516}{105} - \frac{99}{7} x + \frac{11}{21} x^{2} \right)  \right\}  \, ,   \\
         \nonumber  \\
I_{9} & = & \frac{1}{2} \sigma_{T} N_{e} \theta_{e} x^{9} \, \left\{  \,
\theta_{e}^{4} \left(\frac{22}{7} - \frac{11}{42} x \right)   \right\}  \, ,  \\
        \nonumber   \\
I_{10} & = & \frac{1}{2} \sigma_{T} N_{e} \theta_{e} x^{10} \, \left\{ \, \frac{11}{210} \theta_{e}^{4}  \right\}  \, ,  
\end{eqnarray}
where $\sigma_{T}$ is the Thomson scattering cross section and $N_{e}$ is the electron number density.  The expansion parameter $\theta_{e}$ is defined by
\begin{equation}
\theta_{e} \equiv \frac{k_{B} T_{e}}{mc^{2}}  \, .
\end{equation}
In deriving eqs.\ (2.12) -- (2.21), we have ignored $O(\theta_{e}^{6})$ contributions.
Using eqs.\ (2.12) -- (2.21), one can show that the photon number is conserved order by order in terms of the expansion parameter $\theta_{e}$.

  We now apply the present result of the generalized Kompaneets equation to the Sunyaev-Zel'dovich effect for clusters of galaxies.  We assume the initial photon distribution of the CMB radiation to be Planckian with temperature $T_{0}$:
\begin{equation}
n_{0} (X) \, = \, \frac{1}{e^{X} - 1} \, , 
\end{equation}
where
\begin{equation}
X \, \equiv \, \frac{\omega}{k_{B} T_{0}}  \, .
\end{equation}
Substituting eq.\ (2.23) and eqs.\ (2.12) -- (2.21) into eq.\ (2.10), and assuming
$T_{0}/T_{e} \ll 1$, one obtains the following expression for the fractional distortion of the photon spectrum:
\begin{eqnarray}
\frac{\Delta n(X)}{n_{0}(X)} & = & \frac{y \, \theta_{e} X e^{X}}{e^{X}-1} \, \left[  \,
Y_{0} \, + \, \theta_{e} Y_{1} \, + \, \theta_{e}^{2} Y_{2} \, + \,  \theta_{e}^{3} Y_{3} \, + \,  \theta_{e}^{4} Y_{4} \, \right]  \, , \\
    \nonumber  \\
Y_{0} & = & - 4 \, + \tilde{X}  \,  , \\
Y_{1} & = & - 10 + \frac{47}{2} \tilde{X} - \frac{42}{5} \tilde{X}^{2} + \frac{7}{10} \tilde{X}^{3}  \, + \, \tilde{S}^{2} \left( - \frac{21}{5} + \frac{7}{5} \tilde{X} \right) \,  ,  \\
Y_{2} & = & - \frac{15}{2} + \frac{1023}{8} \tilde{X} - \frac{868}{5} \tilde{X}^{2} + \frac{329}{5} \tilde{X}^{3} - \frac{44}{5} \tilde{X}^{4} + \frac{11}{30} \tilde{X}^{5}  \nonumber \\ 
& + & \tilde{S}^{2} \left( - \frac{434}{5} + \frac{658}{5} \tilde{X}  - \frac{242}{5}  \tilde{X}^{2} + \frac{143}{30} \tilde{X}^{3} \right)  \nonumber  \\
& + &  \tilde{S}^{4} \left( - \frac{44}{5} + \frac{187}{60} \tilde{X} \right) \, , \\
Y_{3} & = & \frac{15}{2} + \frac{2505}{8} \tilde{X} - \frac{7098}{5} \tilde{X}^{2} + \frac{14253}{10} \tilde{X}^{3} - \frac{18594}{35} \tilde{X}^{4}   \nonumber  \\
& + &  \frac{12059}{140} \tilde{X}^{5} - \frac{128}{21} \tilde{X}^{6} + \frac{16}{105} \tilde{X}^{7} \nonumber \\ 
& + & \tilde{S}^{2} \left( - \frac{7098}{10} + \frac{14253}{5} \tilde{X} - \frac{102267}{35}  \tilde{X}^{2} + \frac{156767}{140} \tilde{X}^{3} - \frac{1216}{7}  \tilde{X}^{4} + \frac{64}{7} \tilde{X}^{5} \right)  \nonumber  \\
& + &  \tilde{S}^{4} \left( - \frac{18594}{35} + \frac{205003}{280} \tilde{X} - \frac{1920}{7}  \tilde{X}^{2} + \frac{1024}{35} \tilde{X}^{3} \right) \nonumber  \\
& + &  \tilde{S}^{6} \left( - \frac{544}{21} + \frac{992}{105} \tilde{X} \right) \, , \\
Y_{4} & = & - \frac{135}{32} + \frac{30375}{128} \tilde{X} - \frac{62391}{10} \tilde{X}^{2} + \frac{614727}{40} \tilde{X}^{3} - \frac{124389}{10} \tilde{X}^{4}   \nonumber  \\
& + &  \frac{355703}{80} \tilde{X}^{5} - \frac{16568}{21} \tilde{X}^{6} + \frac{7516}{105} \tilde{X}^{7} - \frac{22}{7} \tilde{X}^{8} + \frac{11}{210} \tilde{X}^{9} \nonumber \\ 
& + & \tilde{S}^{2} \left( - \frac{62391}{20} + \frac{614727}{20} \tilde{X} - \frac{1368279}{20} \tilde{X}^{2} + \frac{4624139}{80} \tilde{X}^{3} - \frac{157396}{7}  \tilde{X}^{4}  \right. \nonumber  \\
&  & \, \, \, \, \, + \, \left. \frac{30064}{7} \tilde{X}^{5} - \frac{2717}{7} \tilde{X}^{6} + \frac{2761}{210} \tilde{X}^{7}   \right)  \nonumber  \\
& + &  \tilde{S}^{4} \left( - \frac{124389}{10} + \frac{6046951}{160} \tilde{X} - \frac{248520}{7} \tilde{X}^{2} + \frac{481024}{35} \tilde{X}^{3} - \frac{15972}{7} \tilde{X}^{4}  \right. \nonumber  \\
&  &  \, \, \, \, + \, \left. \frac{18689}{140} \tilde{X}^{5}  \right) \nonumber  \\
& + &  \tilde{S}^{6} \left( - \frac{70414}{21} + \frac{465992}{105} \tilde{X} - \frac{11792}{7} \tilde{X}^{2} + \frac{19778}{105} \tilde{X}^{3} \right) \nonumber  \\
& + &  \tilde{S}^{8} \left( - \frac{682}{7} + \frac{7601}{210} \tilde{X} \right) \, ,
\end{eqnarray}
where
\begin{eqnarray}
y & \equiv & \sigma_{T} \int d \ell N_{e}  \, , \\
\tilde{X} & \equiv &  X \, {\rm coth} \left( \frac{X}{2} \right)  \, , \\
\tilde{S} & \equiv & \frac{X}{ \displaystyle{ {\rm sinh} \left( \frac{X}{2} \right)} }   \, .
\end{eqnarray}
Note that the analytic forms of $Y_{0}$, $Y_{1}$ and $Y_{2}$ in eqs.\ (2.26) -- (2.28) agree with the results obtained by Challinor and Lasenby (1997).
Finally we define the distortion of the spectral intensity as follows:
\begin{equation}
\Delta I \, = \, \frac{X^{3}}{e^{X}-1} \frac{\Delta n(X)}{n_{0}(X)} \, .
\end{equation}

\section{ANALYSIS OF THE CONVERGENCE OF THE POWER SERIES}

  We now carefully study the convergence of the analytic expressions of eqs.\ (2.25) and (2.34).  In order to do the task, first of all, 
we integrate eq.\ (2.9) directly by numerical integration.  We confirm that the total photon number is conserved with excellent accuracy ($< 10^{-9}$) in the numerical integration.  We are now ready to compare the present numerical results with those obtained by the analytic expressions of eqs.\ (2.25) and (2.34) for various $X$-$T_{e}$ regions and investigate the accuracy of the analytic expressions.

\subsection{RAYLEIGH--JEANS REGION}

In the Rayleigh--Jeans limit where $X \rightarrow 0$, eq.\ (2.25) is further simplified:
\begin{equation}
\frac{\Delta n(X)}{n_{0}(X)} \, \longrightarrow \, - 2 y \, \theta_{e} \, \left[ \, 1 - \frac{17}{10} \theta_{e} + \frac{123}{40} \theta_{e}^{2} - \frac{1989}{280} \theta_{e}^{3}
 + \frac{14403}{640} \theta_{e}^{4} \, \right]  \, .
\end{equation}
As is seen explicitly from eq.\ (3.1), the convergence of the power expansion is very fast in the $X \rightarrow 0$ limit.  Furthermore, we  show in Fig.\ 1 the $T_{e}$-dependence of the spectral intensity distortion eq.\ (2.34) for $X=1$.  As is expected, the convergence is extremely fast for $k_{B}T_{e} \leq 50{\rm keV}$.  Relativistic corrections higher than $O(\theta_{e}^{3})$ terms are almost negligible in this region.  So far the Sunyaev-Zel'dovich effects have been measured in the Rayleigh--Jeans region.  Therefore one can reliably apply the analytic expressions of eqs.\ (2.25), (2.34) and (3.1) to the analysis of the observed data.

  In passing we remark that the form of eq.\ (3.1) is meaningful only in an idealized situation.  In order that the higher order terms have a  physical meaning it is necessary that the electron distribution is  rigorously given by the relativistic Maxwellian distribution eq.\ (2.6) with a precisely-determined temperature $T_{e}$. In the real observation, the electron temperature $T_{e}$ has a significant amount of observational error.  This thereby restricts the precision of the 
formula (3.1).

\subsection{$X \approx 4$ REGION}

As one can see from eq.\ (2.26), the leading order contribution $Y_{0}$ vanishes at $\tilde{X}$ = 4 $\approx X$.  Therefore, higher order corrections become more important in this region.  In Fig.\ 2 we have plotted the $T_{e}$-dependence of the fractional spectral distortion at $X = 4$.  It is seen that the dash-dotted line is closest to the exact result.  It should be emphasized here that the dash-dotted line is the contribution including only first four terms in eq.\ (2.25).  The dashed curve which includes all terms in eq.\ (2.25) shows a poorer agreement with the exact result.  This means that the power series expansion in $\theta_{e}$ in this region is not convergent but asymptotic for large $T_{e}$.  We conclude that the analytic expression which includes up to $O(\theta_{e}^{5})$ terms is reliable for $k_{B}T_{e} \leq 15{\rm keV}$ in the $X \approx 4$ region.  We recommend that the analytic expression which includes up to $\theta_{e}^{3} Y_{3}$ terms (dash-dotted curve) be used for the analysis of the observational data for 15keV $< k_{B}T_{e} <$ 30keV, $X \approx 4$.

\subsection{WIEN REGION}

We now study the Wien region where $X > 4$.  As is mentioned earlier, the 
Sunyaev-Zel'dovich effects have been so far studied observationally in the $X \ll 1$ region.  However, the effects will be observed in the Wien region in the future.  For an illustrative purpose, we show the $T_{e}$-dependence of the spectral intensity distortion of eq.\ (2.34) at $X = 8$ in Fig.\ 3.  The convergence is very slow.  All curves are diverging quickly from the solid curve (exact result) for $k_{B}T_{e} > 30{\rm keV}$.  We conclude that the analytic expression including up to $O(\theta_{e}^{5})$ terms is reliable for $k_{B}T_{e} \leq 15{\rm keV}$.  In Figs.\ 4--5 we show $\Delta I/y$ for $k_{B}T_{e}$ = 15keV and $k_{B}T_{e}$ = 20keV, respectively.  We confirm the good accuracy of the analytic expression for $k_{B}T_{e}$ = 15keV.

\subsection{CROSSOVER FREQUENCY}

Finally we study the crossover frequency $X_{0}$, where the spectral intensity distortion vanishes.  It is known that the accurate determination of the $X_{0}$ values is extremely important for the study of the Sunyaev-Zel'dovich effects (Rephaeli 1995).  In Fig.\ 6, we have plotted the $T_{e}$-dependence of $X_{0}$ for $k_{B}T_{e} \leq 50{\rm keV}$ calculated by the analytic expressions and also by the numerical integration (solid curve).  The numerical result is well approximated as a linear function of $\theta_{e}$ for $k_{B}T_{e} < 20{\rm keV}$.  It starts to deviate from the linear form for $k_{B}T_{e} > 20{\rm keV}$.  We have fitted the numerical result as follows:
\begin{equation}
X_{0} \, \approx \, 3.830 \, \left( \, 1 + \, 1.1674 \theta_{e} \, - \, 0.8533 \theta_{e}^{2} \, \right)  \, .
\end{equation}
The errors of this fitting function are less than $1 \times 10^{-3}$ for $0 \leq k_{B}T_{e} \leq 50{\rm keV}$.

  Rephaeli (1995) discusses the consequence of the relativistic shift of the crossover frequency on the value of the peculiar velocity of the cluster measured with the use of the kinematic Sunyaev-Zel'dovich 
effect.  For a cluster with $k_{B}T_{e}$ = 13.8keV, one obtains an error of 20 km/s in the deduced value of the peculiar velocity for the SUZIE experiment corresponding to an accuracy of $1 \times 10^{-3}$ in eq.\ (3.2), by making an interpolation of Rephaeli's estimation.  Therefore, one concludes that the current level of the observational accuracy has not reached the accuracy provided by the theoretical formula (3.2).

  Another impliaction of the accuracy $1 \times 10^{-3}$ of eq.\ (3.2) will be the following.  Let us consider a cluster with the plasma temperature  $k_{B} T_{e}$ = 10 keV making a proper motion with the velocity $v$ = 1000 km/s.  Then the effect of this proper motion on the Sunyaev-Zel'dovich effect will be of order 
$(v/c)^{2}/(k_{B} T_{e}/mc^{2})$ = $5 \times 10^{-4}$.  Thus the kinematic effect of the proper motion of the cluster will not exceed the accuracy 
of eq.\ (3.2).

Finally we estimate an error $\delta$ of the power series approximation in the $X-T_{e}$ parameter space.  We define
\begin{equation}
\delta \, \equiv \, \mid 1 \, - \, \frac{\Delta I}{\Delta I_{num}} \mid 
  \, , 
\end{equation}
where $\Delta I$ is given by eq.\ (2.34) with eqs.\ (2.25)--(2.33) and $\Delta I_{num}$ is the result of the direct numerical integration of 
eq.\ (2.1).  In Fig.\ 7, we have plotted the results for $\delta = 1\%, 5\%$, and $10\%$.  It should be remarked that the sharp peaks in Fig.\ 7 correspond to the points of $X$ where the analytic curve and the numerical curve are crossing.  This can be easily seen in Fig.\ 5 that the dashed curve (analytic curve) and the solid curve (numerical curve) are crossing at these points of $X$.  It is obvious from Fig.\ 7 that the power series approximation is extremely accurate up to 50keV for the Rayleigh--Jeans region ($X<1.5$).  The approximation has a larger error in the $X \approx 4$ region where the leading order contribution vanishes.  Finally the power series approximation is good up to 15keV for the Wien region.

\section{CONCLUDING REMARKS}

We have calculated the relativistic corrections to the Sunyaev-Zel'dovich effect including terms up to the order $O(\theta_{e}^{5})$.  The results of the obtained analytic expressions have been compared with the result of 
the direct numerical integration.  The extent of the applicability of these analytic expressions has been carefully examined.  It has been shown that the present analytic expressions have an excellent accuracy in the Rayleigh--Jeans region where $X \leq 1$.  There the convergence is extremely fast and the results are reliable at least for $k_{B}T_{e} \leq 50{\rm keV}$.  On the other hand, the applicability of the analytic expressions  becomes to be limited for both cases of the $X \approx 4$ region and the Wien region where $X > 4$.  In these regions, the presently obtained analytic expression which takes into account up to $O(\theta_{e}^{5})$ 
terms is reliably  applicable for $k_{B}T_{e} \leq 15{\rm keV}$.  This is the highest electron temperature of the presently known clusters of galaxies.  Therefore, the present analytic expressions can be reliably applied to the known clusters of galaxies.

\acknowledgements
We thank Dr. A. Lasenby, Dr. A. Challinor, and  Professor M. Longair for 
sending us a preprint and for stimulating discussions on this subject. 
We also thank our anonymous referee for many valuable comments which helped us tremendously in revising the manuscript.

\newpage


\references{} 
\reference{} Arnaud, K. A., Mushotzky, R. F., Ezawa, H., Fukazawa, Y., Ohashi, T., Bautz, M. W., Crewe, G. B., Gendreau, K. C., Yamashita, K., Kamata, Y., \& Akimoto, F. 1994, ApJ, 436, L67
\reference{} Berestetskii, V. B., Lifshitz, E. M., \& Pitaevskii, L. P. 1982, $Quantum$ $Electrodynamics$ (Oxford: Pergamon)
\reference{} Birkinshaw, M. 1979, MNRAS, 187, 847
\reference{} Birkinshaw, M., \& Hughes, J., P. 1994, ApJ, 420, 33
\reference{} Birkinshaw, M., Hughes, J. P., \& Arnaud, K. A. 1991, ApJ, 379, 466
\reference{} Buchler, J. R., \& Yueh, W. R. 1976, ApJ, 210, 440
\reference{} Cavaliere, A., Danese, L., \& De Zotti, G. 1979, A\&A, 75, 322
\reference{} Challinor, A., \& Lasenby, A., 1997, preprint of Cavendish Laboratory
\reference{} Chandrasekhar, S. 1950, $Radiative$ $Transfer$ (New York: Dover)
\reference{} Gunn, J. E. 1978, in Observational Cosmology, 1, ed. A. Maeder, L. Martinet \& G. Tammann (Sauverny: Geneva Obs.)
\reference{} Herbig, T., Lawrence, C. R., Readhead, A. C. S., \& Gulkus, S. 1995, ApJ, 449, L5
\reference{} Holzapfel, W. L. et al. 1997, ApJ, 480, 449
\reference{} Jones, M. 1995, Astrophys. Lett. Comm., 6, 347
\reference{} Kompaneets, A. S. 1957, Soviet Physics JETP, 4, 730
\reference{} Markevitch, M., Mushotzky, R., Inoue, H., Yamashita, K., Furuzawa, A., \& Tawara, Y. 1996, ApJ, 456, 437
\reference{} Markevitch, M., Yamashita, K., Furuzawa, A., \& Tawara, Y. 1994, ApJ, 436, L71
\reference{} Myers, S. T., Baker, J. E., Readhead, A. C. S., \& Herbig, T. 1995, preprint
\reference{} Rephaeli, Y. 1995, ApJ, 445, 33
\reference{} Rephaeli. Y., \& Yankovitch, D. 1997, ApJ, 481, L55
\reference{} Silk, J. I., \& White, S. D. M. 1978, ApJ, 226, L103
\reference{} Stebbins, A., 1997, preprint astro-ph/9705178
\reference{} Sunyaev, R. A., \& Zel'dovich, Ya. B. 1980, Ann. Rev. Astron. Astrophys., 18, 537
\reference{} Sunyaev, R. A., \& Zel'dovich, Ya. B. 1972, Comm. Ap. Space Sci., 4, 173
\reference{} Weymann, R. 1965, Phys. Fluid, 8, 2112
\reference{} Zel'dovich, Ya. B., \& Sunyaev, R. A. 1969, Astrophys. Space Sci., 4, 301


\newpage
\centerline{\bf \large Figure Captions}

\begin{itemize}

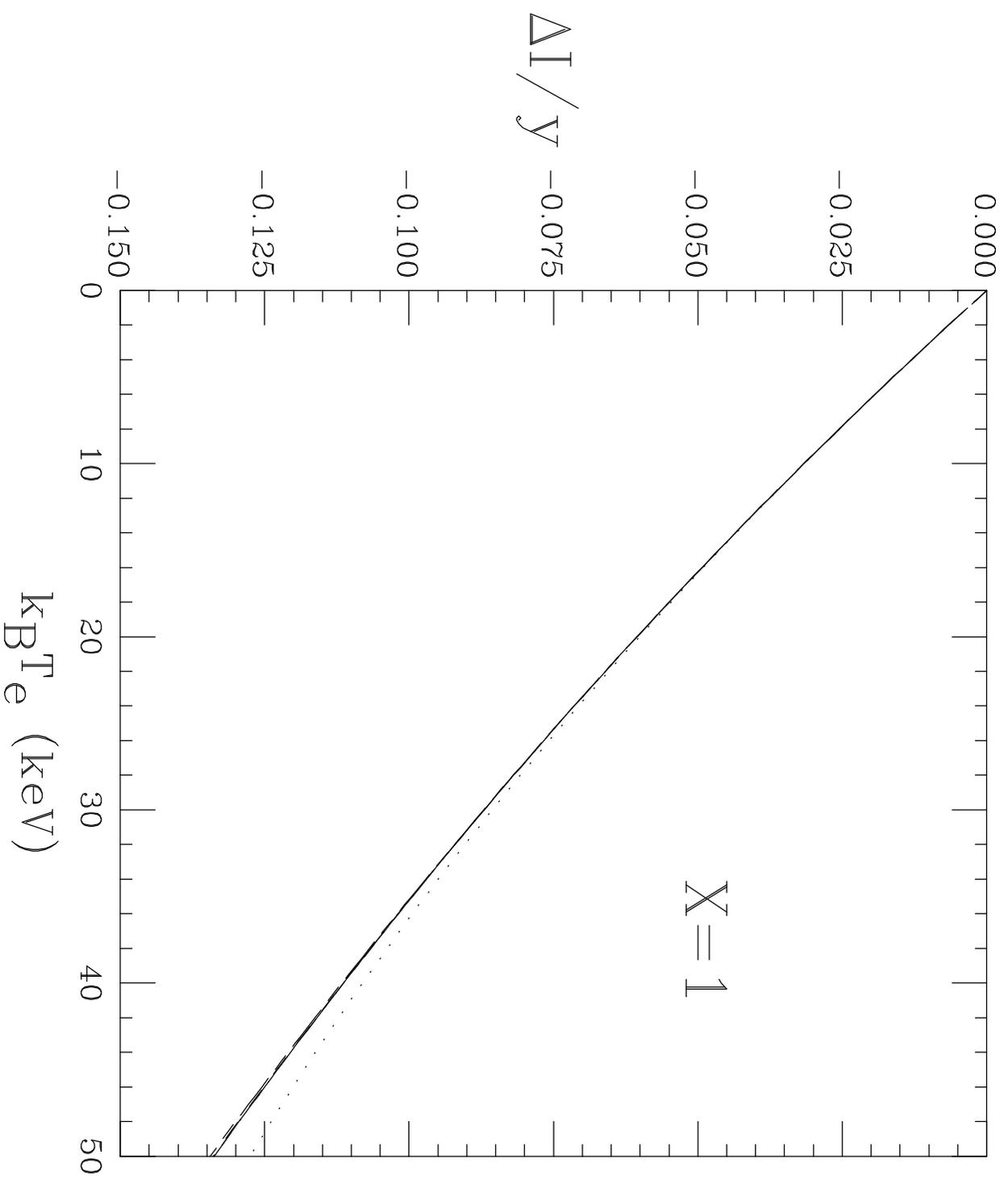
\item Fig.1. Spectral intensity distortion $\Delta I/y$ as a function of $k_{B}T_{e}$ for $X=1$.  The solid curve shows the result of the numerical integration.  The dotted curve shows the contribution of the first two terms in eq.\ (2.25).  The long--short dashed curve shows the contribution of the first three terms.  The dash-dotted curve shows the contribution from the
first four terms.  Finally the dashed curve shows the contribution from all the terms in eq.\ (2.25). 

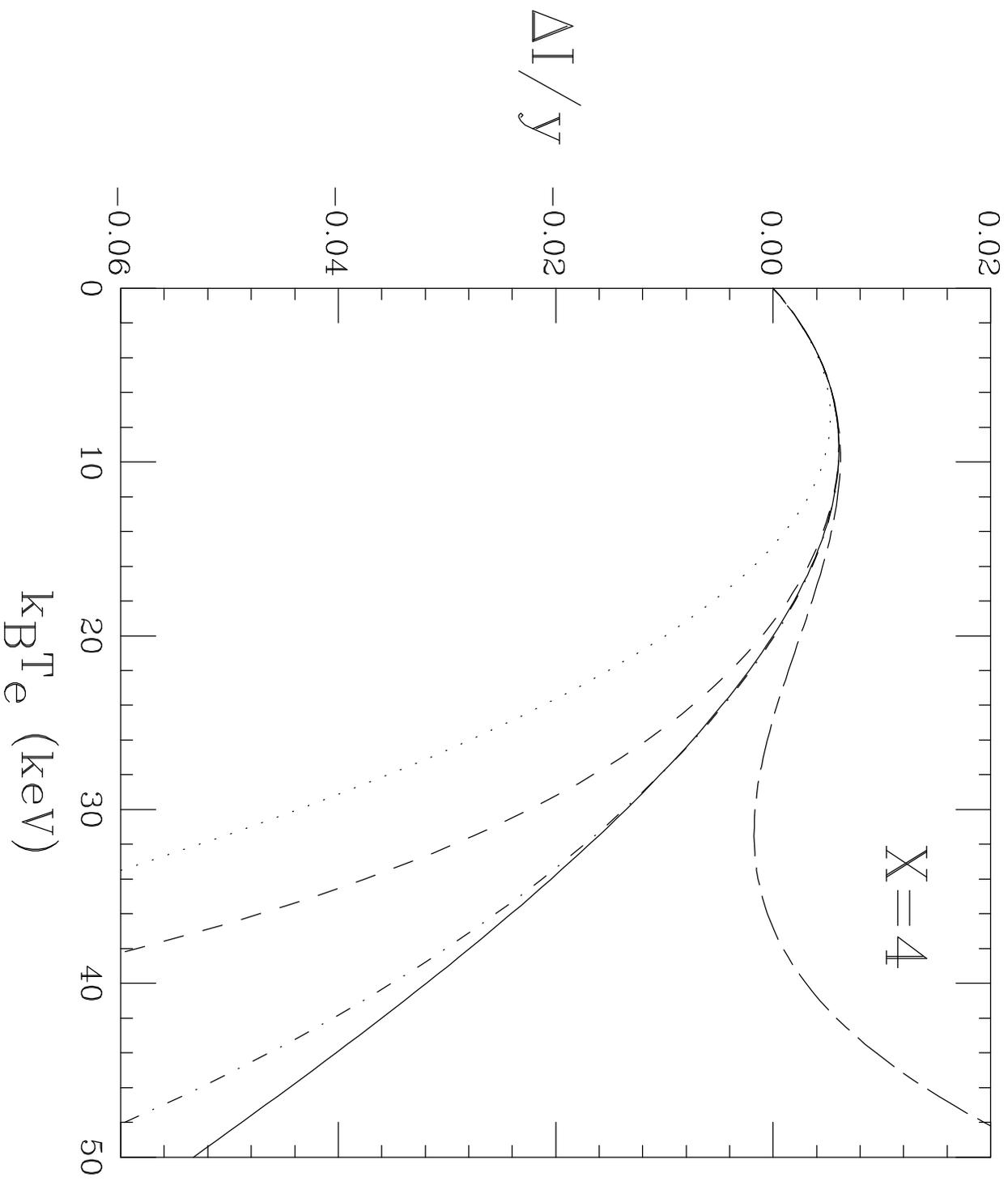
\item Fig.2. Same as for Fig.\ 1 but for $X = 4$. 

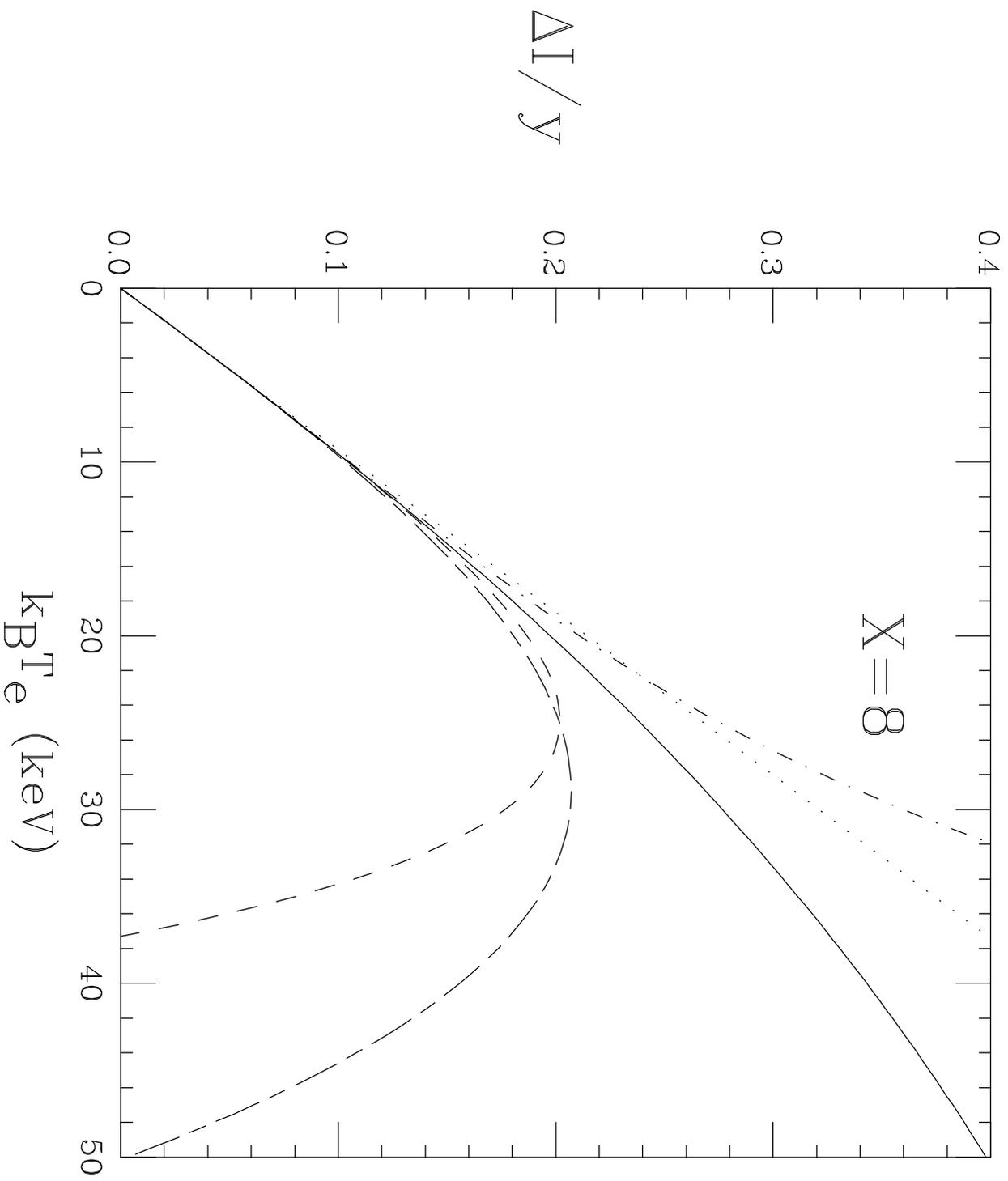
\item Fig.3. Same as for Fig.\ 1 but for $X = 8$. 

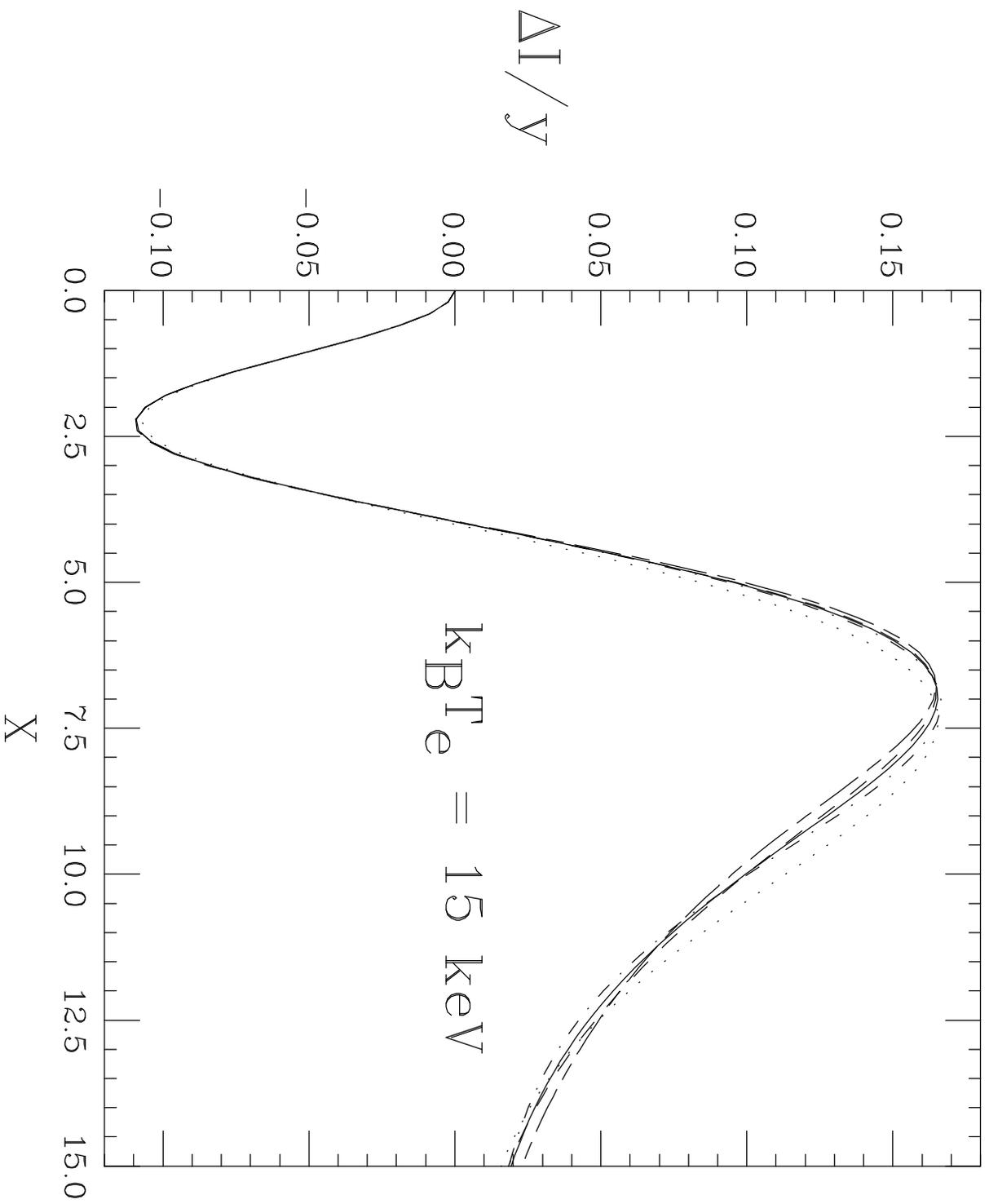
\item Fig.4. Spectral intensity distortion $\Delta I/y$ as a function of $X$ for $k_{B}T_{e}$ = 15keV.  See Fig.\ 1 for the explanation of the curves. 

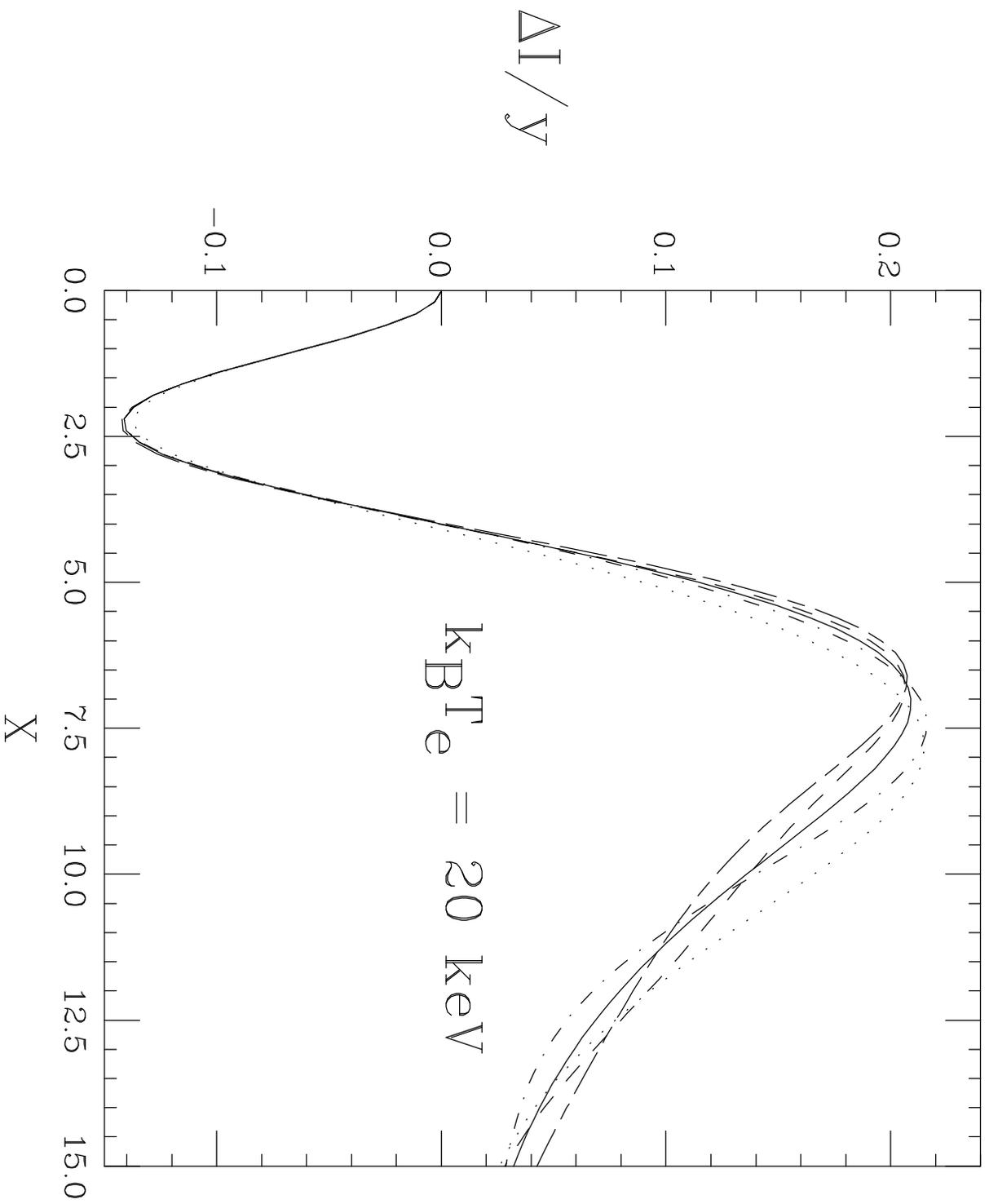
\item Fig.5. Same as for Fig.\ 4 but for $k_{B}T_{e}$ = 20keV. 

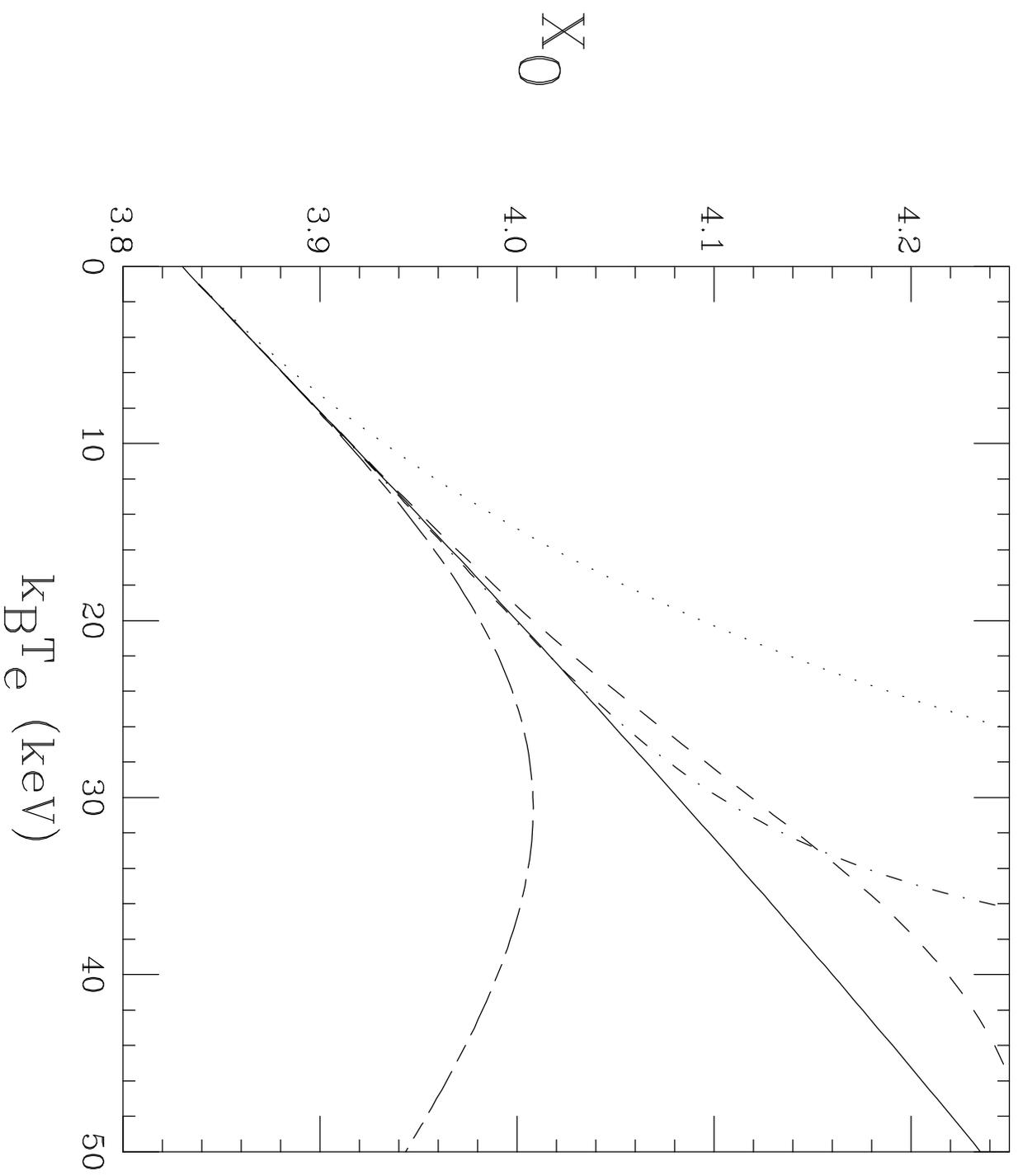
\item Fig.6. Crossover frequency $X_{0}$ as a function of $k_{B}T_{e}$. 
The solid curve shows the result of the numerical integration.  
The dotted curve shows the contribution of the first two terms in eq.\ (2.25).  The long--short dashed curve shows the contribution of the first three terms.  The dash-dotted curve shows the contribution from the first four terms.  Finally the dashed curve shows the contribution from all the terms in eq.\ (2.25).

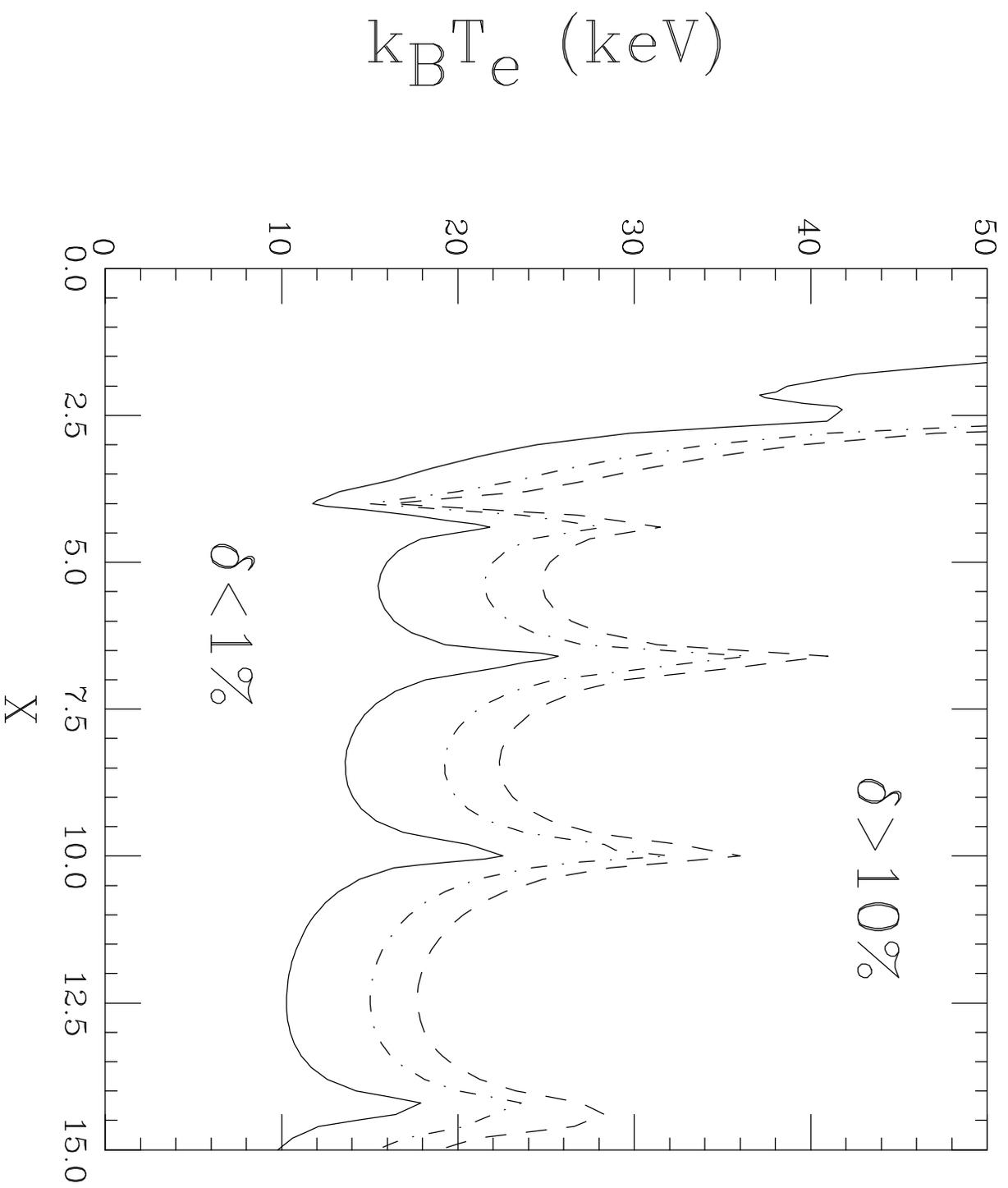
\item Fig.7. Estimation of the error $\delta$.  The solid curve shows the results for $\delta=1\%$.  The dash-dotted curve shows the results for $\delta=5\%$.  Finally the dashed curve shows the results for $\delta=10\%$.

\end{itemize}

\end{document}